\documentclass[12pt,preprint]{aastex}

\begin{document}

\title{The Spiral Host Galaxy of the Double Radio Source 0313-192 
\footnote{Based on observations made with the NASA-ESA 
{\it Hubble Space Telescope} obtained at the Space Telescope Science Institute, 
which is operated by the Association of Universities for Research in Astronomy, 
Inc., under NASA contract No. NAS5-26555; with the {\sl Chandra} X-Ray 
Observatory;
and on observations obtained at the Gemini Observatory, which is operated by 
the Association of Universities for Research in Astronomy, Inc., under a 
cooperative agreement with the NSF on behalf of the Gemini partnership: the 
National Science Foundation (United States), the Particle Physics and Astronomy 
Research Council (United Kingdom), the National Research Council (Canada),
CONICYT (Chile), the Australian Research Council (Australia), CNPq (Brazil), 
and CONICET (Argentina)}}

\author{William C. Keel}
\affil{Department of Physics and Astronomy, University of Alabama, Box 870324,
Tuscaloosa, AL 35487; keel@bildad.astr.ua.edu}

\author{Raymond E. White III}
\affil{Department of Physics and Astronomy, University of Alabama, Box 870324,
Tuscaloosa, AL 35487; rwhite@bama.ua.edu}

\author{Frazer N. Owen}
\affil{National Radio Astronomy Observatory\footnote{The National Radio 
Astronomy Observatory is a facility of the National Science
Foundation, operated under cooperative agreement by Associated Universities,
Inc.}, P. O. Box O, Socorro, NM 87801.
}

\author{Michael J. Ledlow\footnote{Deceased 2004 June 5.}}
\affil{Gemini Observatory, Southern Operations Center, AURA, Inc., Casilla 603,
La Serena, Chile.}

\begin{abstract}
We present new {\it Hubble}, {\it Gemini-S}, and {\sl Chandra} observations
of the radio galaxy 0313-192, which hosts a 350-kpc double source
and jets, even though previous data have suggested that it is
a spiral galaxy. We measure the bulge scale and 
luminosity, radial and vertical profiles of disk starlight, and
consider the distributions of H II regions and absorbing dust.
In each case, the HST data confirm its classification
as an edge-on spiral galaxy, the only such system known to produce such an
extended radio source of this kind. The {\it Gemini} near-IR images 
and {\sl Chandra} spectral fit reveal a strongly obscured central AGN, seen 
through the entire ISM path length of the disk and showing X-ray evidence of
additional absorption from warm or dense material close to the central object. 
We consider several possible
mechanisms for producing such a rare combination of AGN and host 
properties, some combination of which may be at work. These include an
unusually luminous bulge (suggesting a black hole of mass $5-9 \times
10^8$ $M_\sun$), orientation of the jets near the
pole of the gas-rich disk, and some evidence of a weak 
gravitational interaction
which has warped the disk and could have enhanced fuelling of the
central engine. We detect an X-ray counterpart of the kiloparsec-scale
radio jet emerging to the south; jet/counterjet limits on both
radio and X-ray regimes allow them to be symmetric if seen more
than $15^\circ$ from the plane of the sky, still consistent with the
jet axes being within $\sim 30^\circ$ of the poles of the gas-rich galaxy
disk. A linear or disklike emission-line structure is seen around the
nucleus, inclined by $\sim 20^\circ$ to the stellar disk but nearly
perpendicular to the jets; this may represent the aftermath of a
galaxy encounter, in which gas is photoionized by a direct view of the
nuclear continuum. 
\end{abstract}


\section{Introduction}

The relationships between active galactic nuclei (AGN) and their host
galaxies may be important not only for the evolution of the AGN, but for the
host galaxies themselves. Since dynamical evidence supports the existence
of massive black holes in numerous galactic nuclei, and very general
arguments suggest that the active growth of such black holes by accretion
is what we see as AGN, the current demographics of such black holes
must result from the whole history of AGN episodes. Furthermore,
some aspects of galaxy formation become easier to understand if
AGN episodes regulate the properties of gas in galaxies via 
such feedback mechanisms as heating or wind sweeping. These
factors have renewed interest in the connections between properties
of AGN and their host systems.

Among the most durable systematic patterns between the occurrence of AGN
and the structure of surrounding galaxies has been the finding that
large double radio sources occur only around elliptical galaxies,
or systems that plausibly would have been called ellipticals until
a recent external disturbance (in studies ranging from Matthews, 
Morgan, \& Schmidt 1964 to Heckman et al.\ 1986). This correlation has 
figured in
theoretical explanations of large-scale jets, and in understanding
why such jets are associated with extensive synchrotron lobes. 
There are a handful of nearby radio galaxies showing evidence for
gas-rich disks, generally in host galaxies so disturbed that these
are well explained as captured material from major mergers --
3C 120 (Arp 1975; Heckman \& Balick 1979; Garc{\'{\i}}a-Lorenzo et al. 2005), 
3C 293 (van Bregel et al. 1984; Floyd et al. 2006), 
3C 305 (Heckman et al. 1982; Jackson et al. 2003), 
and 3C 285 (Roche \& Eales 2000) 
are well-studied examples. In each of these, the system is
disturbed enough to leave its original galaxy type ambiguous. While
{\it normal} spiral galaxies
with AGN often have kiloparsec-scale radio jets;
no previously reported case of a large-scale double radio source
associated with a spiral has withstood
detailed examination as regards both the host Hubble type and
its identification with the radio source. 
In contrast, the edge-on galaxy associated with the
radio source 0313-192 in the cluster Abell 428 obviously has a significant
stellar disk (Ledlow, Owen, \& Keel 1998), 
and a prominent dust lane of the kind usually associated
with spiral galaxies. We report here {\it Hubble} ACS, {\it Gemini-S},
and {\sl Chandra} ACIS images of
this galaxy, resulting in quantitative photometric evidence that the
galaxy is in fact a spiral, and pointing to peculiar features of its
gas distribution which may shed light on how this rare combination of
galaxy and nuclear activity happened.

As shown by Ledlow et al.\ (2001), the radio structure of 0313-192
includes a nuclear 
source, a nuclear jet toward the southern lobe of about the same
extent as that in M87 but several times less powerful, larger
twin jets directed at projected angles of about 70$^\circ$ to the galaxy 
disk plane, and edge-darkened lobes spanning roughly 280$\arcsec$ (about
360 kpc at $z=0.067$ for ${\rm H}_0$=70 km s$^{-1}$ Mpc$^{-1}$). This 
structure
would typically fall into Fanaroff-Riley type I (FR I), a classification
in which it would lie near the middle of the power distribution
($\approx 10^{24}$ W Hz$^{-1}$ at 1400 MHz). 
Strong 21-cm H I absorption 
is detected against the central source at a column density ${\rm N}_H > 10^{22}$
cm$^{-2}$, with the limit set by its large optical depth and the resolution
of the data 
(Ledlow et al.\ 2001). The H I absorption is also narrow in velocity
span (34 km s$^{-1}$ FWHM), more characteristic of a line of sight cutting
radially through a large disk than one dominated by gas closer to
the central mass. This large column density
is comparable to what we see toward Sgr A, supporting the
idea that we are seeing a galaxy similar to our own viewed exactly
edge-on.

\section{Observations}

\subsection{HST ACS imaging}

Images in broad and narrow bands were obtained with the Advanced Camera
for Surveys (ACS) on board HST, on 16 July 2002, using the Wide-Field Camera 
(WFC). Two broad bands, F555W
($V$) and F775W ($I$), were observed for total exposures of 1100 seconds
each, split for cosmic-ray rejection. The redshifted wavelengths of
[O III] and H$\alpha$+[N II] were observed using the ACS linear ramp filters 
(LRF) for 2500 seconds each, again split for cosmic-ray rejection. 
For [O III] the FR551N ramp filter was centered at 5342 \AA, while
H$\alpha$+[N II] were observed with ramp FR716N set to 7003 \AA.
Since wavelengths are set by rotating the filters in ACS, positioning of
the target was consistent between these two narrow-band images,
requiring only a single set of offsets between continuum and narrow-band images
for registration and continuum subtraction. These filters have nominal
half-power bandwidths of $\Delta \lambda/\lambda= 0.02$. Thus, the [O III] setting
excludes the $\lambda 4959$ line, while the H$\alpha$ image
includes the adjacent [N II] lines.

To verify the registration of the optical and radio coordinate systems,
we used 11 stars and compact galaxies listed in the USNO A-2 astrometric
catalog lying within the ACS field on the broadband images. The 
coordinate offsets derived from individual stars have scatter  $\sigma = 0.1$
arcsecond (2 ACS WFC pixels), so we consider that coordinates derived from
the HST frames using this correction are on the VLA reference frame to
roughly this accuracy. The overall correction from header to astrometric
coordinates using these reference stars
is -0.1016 seconds in RA and -1.26$\arcsec$ in declination.

                                                                                                                                                                                                                                                                                                                                                                                                                                                                                                                                                                                                                                                                                                                                                                                                                                                                                                                                                                                                                                                                                                                                                                                                                                                                                                                                                                                                                                                                                                                                                                                                                                                                                                                                                                                                                                                                                                                                                                                                                                                                   
Based on this registration, the radio core coincides closely with the
galaxy nucleus, as measured from the center of the $I$-band bulge isophotes.
The radio core ($\alpha=$03:15:52.10, $\delta=$-19:06:44.4 in the J2000
frame) is within 0.20" of the
bulge center, as derived from ellipses fit to
large arcs of the isophotes at a semimajor axis
of 1". Such a tight match (within $2 \sigma$) gives us 
confidence that the radio source
and galaxy are indeed the same object. For visual reference, Fig.\ 1
shows a composite VLA A-array image, where the outer regions are 
as seen at 1.4 Ghz and the inner jet is from an 8.5-Ghz observation,
superimposed on the color rendition of the ACS imagery.
The prominent grand-design spiral to the northwest is
not a good candidate for a physical companion, lying at 
$cz=15225 \pm 87$ km s$^{-1}$
(Ledlow et al.\ 2001), as opposed to a redshift of $cz=20143 \pm 66$
for 0313-192 itself. A combination of sparse galaxy population and
large redshift spread led Ledlow et al.\ to suggest that Abell 428
is more like a filament seen end-on, rich in small groups, 
than a cluster environment.

Fig.\ 2 shows the smaller-field view in the continuum, as a composite
color display. This view emphasizes the intricate structure in
the disk dust lane, and the blue star-forming regions seen
around the disk warp.

Continuum-subtracted emission-line images were constructed using the
broadband images as continuum, with the correction for line radiation in
these bands being small. The relative shifts between broad- and narrow-band
images could be checked from the data themselves only to the $\pm 2$ pixel
level, since there are no bright unresolved sources in the region of
useful LRF transmission, but this should cause no particular
problem in the analysis since we see no features with the positive/negative
pattern which would result from subtraction of an inappropriately shifted
continuum. The scaling factors to account for the continuum included
in the narrow bands were empirically determined from field stars and
the galaxy bulge. For H$\alpha$, since the line wavelength is near the blue
edge of the F775W passband, an improved continuum subtraction was
possible using an interpolated 7000-\AA\  image obtained as a weighted
average of the $V$ and $I$ images. The emission-line images are
shown in Fig.\ 3. The color of reddened light in the dust lane is
extreme enough that we cannot simultaneously get a proper continuum
subtraction there and in the rest of the galaxy; some negative residuals 
appear in the dustiest areas.

In a search for UV continuum emission from the inner knots in the radio
jet on the SW side of the nucleus, we also obtained a UV continuum image
using the ACS High-Resolution Camera (HRC) and F250W filter, 
in a 2600-second exposure pair. A few faint sources are detected
near the galaxy but none coincident with either disk star-forming
regions or jet knots. This stands in contrast to our soft X-ray detection
of the southern jet out to about 3$\arcsec$ from the core (section 5).
Our detection limit for compact sources allows a single power-law
specrtum between radio and X-ray measurements.


\subsection{Gemini-S NIR imaging}

Images of 0313-192 in the $J$,$H$, and $K_s$ bands were obtained using the
FLAMINGOS system at the 8-m {\it Gemini-South} telescope, during
a commissioning run on 13 October 2001.
For each, a set of 100-second exposures in a $3 \times 3$-pointing
offset pattern was combined, with offset sky measurements interspersed.
The mean delivered image quality (DIQ) was 0.39" FWHM, sampled with
0.078" pixels. The images
are compared in Fig.\ 4. Their resolution is sufficient to reveal
the dust lane clearly at J and H, while a central source is
progressively more prominent at H and K. This light comes from
some combination of the inner starlight, increasingly visible
through the dust lane, and the AGN itself, possibly even more heavily
reddened. 

We used the Apache Point data from Ledlow et al.\ (1998) to set the
photometric scale for the FLAMINGOS images. The central region,
within the seeing disk, has $J-K = 1.6$, compared to the surrounding
bulge light at $J-K \approx 1.0$. The spatial distribution of the redder
component is consistent with a contribution from a central point
source. This, coupled with the
$K$ magnitude of the central source, suggests that we are
seeing the central AGN in addition to reddened light from the obscured
part of the bulge (section 4).

\subsection{Chandra ACIS observations}

A 19.2-ksec observation of 0313-192 (ObsID number 4874)
was obtained in February 2004,
with the galaxy located at the ACIS-S aimpoint. This was done
for best sensitivity at lower energies, since the active nucleus
was likely to be deeply obscured by foreground gas. This proved
to be the case, with strong low-energy absorption. 
As discussed in section 4, the best fit for an absorbed power-law spectrum
has a large column density $N_{\rm H}=4.06^{+0.73}_{-0.58}\times10^{22}$
cm$^{-2}$ and a hard photon index $\Gamma=1.33^{+0.34}_{-0.27}$.

The strong absorption against the nucleus allows a sensitive search
for surrounding structures at low energy. As discussed in section 5, 
this approach allows
us to detect emission below about 3 keV from the southern jet
previously mapped at 3.6 cm by Ledlow et al. (2001).

\section{Morphology of 0313-192}

The key question we address with the new optical and near-IR 
data is the morphological type
of this galaxy --- is there unequivocal evidence that it is a spiral? 
The ACS images show the classic appearance of an edge-on disk galaxy
(Fig.\ 2). Both the dust lane and stellar disk indicate that we view this galaxy
within $0.5^\circ$ of edge-on, so morphological classification
must rely on secondary characteristics rather than on the properties of
the spiral arms. These images are of high enough resolution that we
can compare 0313-192 in some detail to local galaxies which are
agreed to be edge-on spirals of well-constrained Hubble type.

\subsection{Bulge}

The ACS data allow a much better view of the bulge than previous images,
so we can fit regions free of significant obscuration to derive
unambiguous luminosity and scale values.
The bulge isophotes, traced in regions where dust is not a problem, remain
slightly boxy, which limits the precision with which we can fit
a bulge model. We fitted a variety of de Vaucouleurs ($r^{1/4}$)
profiles to the bulge, varying the effective radius and axial
ratio, and assessed the goodness of fit by RMS residuals in
several apparently dust-free bulge regions, after subtracting
an approximate disk model.
Our best bulge fit to the
$I$-band image has axial
ratio $b/a=0.6$ and effective radius 50 pixels (2.5\arcsec ), which is
likely determined at the 10\% level. Normalizing to fit the 
brightest bulge regions gives this fit a total brightness
corresponding to Cousins $I=15.08$, taking the magnitude transformations
from the STMAG system definition and Fukugita et al.\ (1995). This
translates to a very luminous bulge, brighter than our earlier
estimates (Ledlow et al.\ 1998) mostly because a significant
fraction of the bulge luminosity is obscured by the dust lane. 
The bulge luminosity will be used in section 4 to
consider the expected mass of a central black hole.
 


\subsection{Stellar disk}

Surface photometry of edge-on spirals has been used extensively to study
the vertical structure and dynamics of disk stars, providing a large
body of comparison data. As found originally by
van der Kruit \& Searle (1981), the starlight distribution parallel to
the disk plane in edge-on spirals, after bulge subtraction, is well expressed 
by the projection of a radially truncated exponential disk, with this 
truncation occurring at radii of several exponential scale lengths. 
Similar behavior
is found in 0313-192 (Fig 5). After subtracting our bulge
model, we find a good fit for radial scale length of 109 pixels (5.4\arcsec, 
6.9 kpc) 
truncated at 20.8 kpc (3.0 scale lengths, which projects to 16.2\arcsec). This 
applies 
for a range of height
north and south of the disk plane, once we adjust the assumed position angle
of the disk for best symmetry on east and west sides (determining
this orientation at the $0.1^\circ$ level). The limit in measuring the
orientation and scale length is set by small asymmetries in the disk which
persist on both north and south sides and in both passbands.

Similarly, the profile of edge-on spiral disks perpendicular to the
disk plane is well described by a single exponential, nearly independent
of radius (van der Kruit \& Searle 1981; de Grijs, Peletier, \&
van der Kruit 1997). This also applies to 0313-192 (Fig.\ 5). Studies
of edge-on galaxies with a range of morphological type have shown that
the ratio of disk scale length to vertical scale height varies
systematically with stage along the Hubble sequence, in the sense that
the disk is typically thinner relative to its radial scale length
for later Hubble types (Schwarzkopf \& Dettmar 1997). Using the
data from Pohlen et al.\ (2000), who tabulate scale heights which
are twice the scale height fitted for an exponential, disks as 
thin as we find in 0313-192 are found only for spirals of type Sb
(de Vaucouleurs T type 3) and later (the S0 distribution includes
only a few which are nearly this thin). This is further 
evidence that, were 0313-192
to be viewed more face-on, it would appear as a clear spiral.
Note, in this context, that some analyses use either a theoretically motivated
sech$^2 (z)$ form or a compromise sech $(z)$ law in fitting the
vertical disk profile; the available observations do not clearly
support one of these as superior to the empirical (albeit
theoretically problematic) exponential version.

The $V-I$ colors of the bulge and most of the disk are quite similar.
This fact, and the scale height of the disk in 0313-192, suggest that much of
what we see is a ``thick disk". However, the blue star-forming layer
in the midplane can be seen where the warp allows us to see around
the dust lane, and its intensity profile forms a virtually continuous
exponential with the starlight farther from the plane. Thus, the
parts of the disk we can see may form a useful basis for estimating the
bulge:disk flux ratio, interpolating across obscured regions
with the combination of de Vaucouleurs
bulge model and double-exponential disk. Our best-fit parameters yield
an $I$-band bulge:disk ratio of 0.4; this would be slightly smaller
in the $V$ band because the midplane star-forming regions become
more important. Again, from our vantage point, both components suffer
strong optical obscuration, so this estimate of a dust-free value
differs from our initial estimates made directly from lower-resolution images.
Even neglecting hidden star-forming regions, our photometric model
makes 0313-192 a very luminous spiral; only 10 spirals of the 895 in the Revised
Shapley-Ames Catalog (Sandage \& Tammann 1981) would outshine it, and
such an apparent-magnitude-limited sample is already biased on favor
of luminous galaxies.


\subsection{Star-forming regions}

The [O III] and H$\alpha$+[N II] images (Fig.\ 3) show discrete objects in the
disk plane which are bluer than their surroundings in $V-I$. These
are normal H II regions, as shown by the emission-line ratios
spatially integrated over large regions of the disk (Ledlow et al.\ 1998)
We see these at optical wavelengths only thanks to the warped disk plane,
which allows large windows where our view penetrates close to the midplane.
 
The only ambiguity in the galaxy classification is whether 0313-192 could 
possibly be an S0. When S0s host significant star formation, as traced via
H$\alpha$ imaging, its distribution is sharply confined on the
outer edge, in an annulus with or without a filled center of star formation
(Pogge \& Eskridge 1993, Macchetto et al.\ 1996,
Koopmann \& Kenney 2006). This distribution follows the locations
of their dust lanes. In 0313-192, star-forming regions seen
around the dusk warp extend as far as the dust, and both extend
as far as 13$\arcsec$ (17 kpc) from the nucleus. The stellar disk can be
traced only slightly farther, to about 21 kpc.

\subsection{Dust lane}

The dust lane shows a distinct warp, by about $3^\circ$ in projection
with respect to the midplane of the stellar disk. This appearance could
result even if both share an identical warp, because of the weighting of
dust visibility to the front side of the disk while the starlight 
as seen around the dust is
closer to the line-of-sight integral of its distribution.

The dust lane in 0313-192 is thick, exhibits rich vertical structure, and
spans virtually the entire radial extent of the stellar disk. All these
features are characteristic of spirals with active disk star formation,
rather than of S0 galaxies which may also have dust in the disk plane.
As shown in detail in Fig.\ 6, the thick opaque dust lane is accompanied by
localized features extending to 0.8 kpc from the plane. These are often
seen in nearby edge-on spirals, with some evidence that their occurrence
and extent correlate with the disk star-formation rate (Howk \& Savage 1999).
Such lanes are generally much thinner in edgewise S0 systems, although
recent Hubble Heritage imaging of NGC 5866 shows that even these galaxies
can show a detectable level of vertical stirring in their dust lanes.

As noted above, the dust lane is detected to a radial distance of 
12.5$\arcsec$ from the nucleus (projecting to 16 kpc), nearly
the same extent over which the stellar disk is found. This also argues
for a genuinely spiral nature, since S0 galaxies with dust lanes
show them over only a fraction of this radial extent. In every instance
of edge-on S0 systems shown in the Carnegie Atlas (Sandage \& Bedke 1994),
and as discussed in their text accompanying images of dusty S0 galaxies,
the dust is in an annular distribution truncated well within the visible
stellar disk, which is not necessarily the case in spirals. This is
true for S0 galaxies classified not only in the revised Hubble
system developed by Sandage, but with the slightly different
criteria used in the de Vaucouleurs system (Buta, Corwin, \& Odewahn
2006). In both classification systems, edge-on spirals have dust distributions
which are less confined radially, and can have the same scale length
and detected extent as the stellar disk. 
Thus, the dust structure in 0313-192 is most clearly characteristic of 
a spiral galaxy rather than even a relatively dust-rich S0 system.
We illustrate this through a comparison (Fig.\ 6) of 0313-192 with
several nearby galaxies having widely-agreed-upon Hubble types. These
local examples are all bright, with large angular size, and
have concordant Hubble types in both the Sandage and de Vaucouleurs
extensions of the Hubble system (using types taken from the Sandage \& Tammann
1987,
Sandage \& Bedke 1994, and Buta et al.\ 2006). This comparison is
in the spirit of the demonstration by Curtis (1918) that spirals
seen at various angles form a continuous sequence, and that their disk planes
are marked by optically thick absorbing regions. In fact, all of the
nearby comparison galaxies we show were also included in Plate III
of Curtis' discussion.


\section{The nucleus}

The reason for a spiral galaxy producing such a rare manifestation of
nuclear activity as this double radio source might lie ultimately
in unusual properties of the AGN itself. Previous data have
illustrated the existence of a core radio source and emission-line
gas characteristic of type 2 Seyferts or narrow-line radio galaxies.
Our new data allow us to probe additional properties of the nucleus
in several ways.

{\subsection{A tilted circumnuclear emission-line structure}

The emission-line images (Fig.\ 3) show a roughly linear structure
inclined by about $20^\circ$ to the disk plane. 
This structure has a higher excitation level than the emission-line
gas seen elsewhere in 0313-192. Using the nominal filter throughput 
values from the ACS online exposure calculator, the disk H II regions
have [O III] $\lambda$5007/(H$\alpha$+[N II] ratios in the range
0.39-0.48, typical of H II regions in luminous spirals with metal-rich
disk gas. The anomalous tilted emission regions have consistent
values on both sides of the dust lane, 
[O III] $\lambda$5007/(H$\alpha$+[N II] = $1.00 \pm 0.03$. A slightly
lower value (0.85) is seen for gas near the nucleus north of the dust lane
away from the tilted structure, and the prominent single knot 
southeast of the nucleus has a level (1.02) consistent with
the anomalous feature. This emission must have been mostly responsible 
for the Sy 2-like spectroscopic signature found by Ledlow et al.\ (1998).
The only significant ionization structure within this feature
is that the [O III] emission is relatively weaker near the edge of
the dust lane, which can be plausibly attributed to extinction.

These data give us very limited information on the nature
of this structure, and even whether it might be part of an inclined
disk of gas or more akin to an ionization cone. The latter seems
unlikely because it makes a large angle with the inner radio jet axis
($\approx 80^\circ$ in projection), and the radio jet falls outside the 
projection of the putative ionization cone. An inclined gaseous
disk would fit with the warped stellar disk as evidence of
a minor merger or more distant encounter, perhaps several disk-crossing
times before our current view. In this view, this gas, and a single knot 
(to the upper right of the nucleus in Fig.\ 3) some 
distance from the main gaseous disk, must be ionized
by a direct view of the central AGN, to account for their similar
excitation levels.

If it is part of a ring or inclined disk, it is noteworthy that
this gas distribution is within $10^\circ$ in projection of
being perpendicular to the radio axis, which would make sense if
the central black hole were accreting material with this
same overall direction of angular momentum. Since we see warps
in two scales -- the warped dust disk
(which may itself sample a warp which exists in the stellar disk,
but is less apparent due to the different weighting of starlight
along the line of sight), and the inclined potential inner
gas-rich disk -- both might be attributed to a minor merger or
other gravitational disturbance. While minor warping can in principle
be excited by internal mechanisms (Bertin \& Mark 1980), the
$20^\circ$ tilt of the emission-line feature seems to require
an externally forced origin. Such a disturbance might increase
the accretion rate of the central black hole and be a factor in
producing this unique manifestation of activity from a spiral galaxy.
However, the timescales for differential precession should scale
with orbital period, meaning that the inner structure should be
more phase-wrapped and less well-defined than the outer warped structure
(as in the evolution of an accreted polar ring as well
as less dramatic tidally-induced warps; Hunter \& Toomre 1969). In 
this light,
producing both warps with a single event would require either that
there are two narrow annuli, so that differential precession could
leave each annulus relatively intact, or the two warps have different
origins. For example, the dust warp might be due to tidal forces
during an encounter from which the disturber was mostly accreted,
with its gas forming the inner emission-line structure.

\subsection{IR detection of a reddened AGN}
The central very red structure in the {\it Gemini} images (section 2.2)
is well explained by a reddened point source located behind the dust
lane. However, it is not bright enough to absolutely require this
to be an AGN rather than the reddened peak of the bulge starlight.
We experimented with subtracted scaled versions of the best-fit
de Vaucouleurs bulge (section 3.1) from the $K_s$ image, and find that
the central intensity always falls short of the unreddened inward
extrapolation of the bulge light. A tighter constraint incorporates the
$J-K$ color excess from the dust lane, which is just resolved in the
{\it Gemini} images. Relative to the adjacent unobscured disk light,
the dust lane is 0.7 magnitude redder in $J-K$. For a Galactic
reddening law, $E(J-K) = 1.8 A_K = 0.18 A_V$. In this case, these values
suggests that material directly behind a typical piece of the dust
lane is obscured by only 1.3 magnitude at $K$. The central flux shortfall
of the scaled bulge model (when matched to the outer bulge isophotes,
yielding disk residual isophotes which are nearly straight across 
the bulge region) is almost exactly a factor 2, implying that the galaxy
has a central flux excess of 64\% of the central bulge luminosity
(within a radius of 0.94$\arcsec$), or roughly $K=15.7$ after correcting 
for our mean extinction in the dust lane. Structure in the dust lane
could change this limit; in particular, it is likely to be
an underestimate given the nondetection of a central source in
even the $I$ ACS image. Taken as a lower limit, this implies that
the AGN has $M_K < -21.5$. Typical QSOs are brighter than
$M_K= -23.5$ (e.g., Percival et al.\ 2001).

\subsection{Hard X-ray continuum}

X-rays may give us the most direct view of the nucleus in this
system, given the ambuguity in unravelling the radio contributions
of the core and small-scale jet emission.

We used {\sl XSPEC} to fit an absorbed power law
to the {\sl Chandra} spectrum, extracted within a 2$\arcsec$ radius.
We adopted a fixed Galactic column of $3.1\times10^{20}$ cm$^{-2}$
(derived from the {\sl HEASARC} nH tool) and included
an intrinsic absorption component at the redshift of the galaxy.
With a count rate of $\sim 0.1$ cnt s$^{-1}$, the source
is mildly piled up, so we applied the {\sl XSPEC} pileup model.
The best fit (Fig. 7) has a $\chi^2$ per degree of freedom
$\chi^2_\nu=1.11$, a photon index of $\Gamma=1.33^{+0.34}_{-0.27}$
(90\% confidence intervals)
and an intrinsic absorbing column of
$N_H=4.06^{+0.73}_{-0.58}\times10^{22}$ cm$^{-2}$.
This absorption is broadly comparable to the $10^{23}$ cm$^{-2}$
reported toward Sgr A itself (Baganoff et al.\ 2003), which make sense
since both lines of sight pass through much of the disk plane of
spiral galaxies.
The unabsorbed flux for this model fit would be
$3.34\times10^{-12}$ erg s$^{-1}$ cm$^{-2}$ in the 0.5-4.5 keV band
and $4.88\times10^{-12}$ erg s$^{-1}$ cm$^{-2}$ in the 2-10 keV band.
These translate into luminosities of $L_X=3.6\times10^{43}$ erg s$^{-1}$
(0.5-4.5 keV) and $L_X=5.3\times10^{43}$ erg s$^{-1}$ (2-10 keV).
These luminosities are
comfortably in the range occupied by both broad- and narrow-lined
radio galaxies (e.g. Grandi et al.\ 2006) and only slightly less luminous 
than recent detections of type II QSOs (Ptak et al.\ 2006). 

We considered a partial-covering
fit to the X-ray data to (at least formally) improve the fit at the 
lowest energies. Including a variable covering factor does
slightly improve the $\chi^2$, by $\Delta \chi^2 = 6.6$ on
adding one additional parameter (to $\chi^2 = 118$ for
111 degrees of freedom). The absorbing column density
increases by 6\% with respect to the range listed anove, for
complete coverage, and the photon index steepens by 0.13 (both changes
lying well within the confidence intervals). The fitted
covering fraction is essentially unity, with the fit yielding
a value of 0.995 but raising the implied column density by $\sim 20$\%.

As set out by Evans et al.\ (2006), there has been considerable discussion
about whether the X-ray emission from radio-galaxy cores arises from
the central engine itself or from a parsec-scale jet. Evans et al.\ 
consider the ranges of X-ray luminosity, absorbing column density,
and spectral shape over a variety of bands to address the issue,
noting that the mean X-ray properties are quite different between 
FR I and FR II sources. In their sample, FR I
radio galaxies have modest absorption, with only one of 15 having
$N_H > 10^{23}$ cm$^{-2}$, while all but one of the 7 FR II nuclei
have column densities above this value. They attribute this to
dense circumnuclear tori in FR II nuclei. In the case of 0313-192,
the geometry of the host galaxy combined with the amount of
foreground absorption mean that we must consider a substantial role
for absorption in the host galaxy as well as around the nucleus.

The equivalent column density found from X-ray absorption is
$N_H= 4 \times 10^{22}$ cm$^{-2}$ (above). This is significantly
greater than the H I column density measured from 21 cm absorption
(${\rm N}_H \approx 10^{22}$ cm$^{-2}$; Ledlow et al. 2001). The
conversion from absorption profile to column density includes
the H I spin temperature $T_s$ and the ionized fraction of
hydrogen, so that absorption by progressively warmer and more
ionized gas leaves a smaller H I signature.  
This mismatch leads us to consider more complex models for
the absorbing gas. Much of the H I is undoubtedly in normal cold
clouds in the galaxy disk, but conditions near the AGN could differ
in several ways. In dense regions, $T_s$ approaches the kinetic temperature,
decreasing the absorption per atom (for example, by a factor $\approx 4$
at $T_s=600$ K). Similarly, fully ionized absorbing gas appears only
in the X-ray accounting. Just as X-ray absorption in excess of the
foreground Galactic H I value is often considered to be intrinsic
to active nuclei, we may here be dealing with absorption from warm
or dense material close to the active nucleus accounting for more
than half the X-ray value. Since we probably view the inner jets
within $30^\circ$ of the plane of the sky, any dense gaseous
torus would be viewed within $30^\circ$ of edge-on. Polar opening
angles of these structures are typically no smaller than this,
to account for the relative numbers of types 1 and 2 Seyfert galaxies
and broad- and narrow-line radio galaxies as well as structures of
ionization and scattering cones (e.g., Antonucci 1993, Clarke et al. 1998), 
so it is plausible that
this excess absorption arises in such a structure very close to the
central source. The circumnuclear absorption may be strong enough to
require an unusually extensive or dense torus; both host-galaxy and
derived circumnuclear absorption are an order of magnitude larger
that in the Seyfert nucleus IC 4329A (Steenbrugge et al. 2005), 
whose host galaxy is viewed with similar geometry. Analogous evidence for
warm or dense circumnuclear H I in Seyfert galaxies has been discussed by
Gallimore et al. (1999), particularly with regard to suppression of 
H I absorption near the active nuclei. This warm or dense absorber
might also contribute to the spectral turnover observed between 1.4 and 8.5 Ghz
in the core of 0313-192 (Ledlow et al. 2001), through free-free absorption,
although the density dependence of the emission measure needed to
calculate its strength makes comparison with column densities
difficult.


\subsection{Bulge properties and the expected black-hole mass}

The bulge luminosity-black hole mass relation, while showing much
more scatter than the relation between central mass and stellar
velocity dispersion, can offer a hint as to the nature of the
central object. Using $K$-corrections and colors for old stellar populations
from Kinney et al.\ (1996), the bulge magnitude $I=15.08$ corresponds to
$M_B=-21.2$. This is quite luminous for a spiral bulge, at about 
$5 \times 10^{10}$ $L_\sun$, and hence with a stellar
mass in the bulge of about $3 \times 10^{11}$ $M_\sun$. 
For comparison, this is about 10\% more luminous than the
bulge of the luminous and bulge-dominated spiral 
M104 (e.g. H\"aring \& Rix 2006), and roughly a full magnitude more
luminous than any of the 86 bulges observed by de Jong 1996 when
transformed to our distance scale.
Surface photometry of M104 from Jarvis \& Freeman (1985) and
Wainscoat et al.(1990) shows that the bulge effective radii are
comparable, to within uncertainities in the relative distances
caused by the Virgocentric velocity field.

The black hole mass - bulge luminosity fit presented 
for the highest-quality sample of
Ferrarese \& Merritt (2000) implies a central mass $8 \times 10^8$
$M_\sun$ for this luminosity; the fit for their ``less certain" sample differs
substantially in this range, giving $3 \times 10^9$, which they attributed
largely to systematics in modelling central masses when the sphere of
influence is not well resolved in the dynamical measurements. 
Indeed, the reconsideration by H\"aring \& Rix suggests a black-hole
mass of $5 \times 10^8$ $M_\sun$ for such a bulge. 


\section{An X-ray jet from a spiral galaxy}

The strong absorption against the core X-ray source allows us
to look for small-scale extended components at low energies
(as also works, for example, for Centaurus A; Kraft et al.\ 2002).
In this case, the core is so faint at low energies that the ACIS
data reveal a jet on even smaller angular scales than those
recently observed in numerous quasars.
As shown in Fig.\ 8, the structure of 0313-192 changes
with energy in a striking way. At the softest energies,
we see a jet about 3$\arcsec$ (4 kpc) long, matching the kpc-scale radio
jet mapped by Ledlow et al.\ (2001). Including photons
up to about 1.5 keV, the jet is surrounded by a more extensive
sructure which remains too one-sided to attribute to the core
PSF. At the highest energies, the structure is completely dominated
by the core.

This X-ray jet matches the inner radio jet as well as our resolution
and limited photon statistics can show (Fig.\ 9). We derive an
approximate flux from 0.5-3 keV by using an absorbed  power-law spectrum of
index $\Gamma=1$. The jet flux is
$\approx 8.5 \times 10^{-15}$ erg cm$^{-2}$ s$^{-1}$, corresponding to
a luminosity of $9.3 \times 10^{40}$ erg s$^{-1}$ in the same band.

This jet is comparable to the jet in M87 in its X-ray properties,
as well as the limited photon statistics will let us say. The
jet of M87, excluding nuclear emission, has a nearly identical
luminosity: about $9 \times 10^{40}$ erg s$^{-1}$ in the same
0.5-3 keV band, using the knot fluxes from Marshall et al.
(2002) and Perlman \& Wilson (2005). We averaged the two sets of
data; more detailed treatment is unwarranted given that the
X-ray flux is variable due (at least) to an occasionally very
bright inner knot (i.e. Harris et al. 2006). The radio flux from
this part of the jet implies a luminosity rather fainter than in
M87, by about a factor 3 (using the 20-cm flux data from Sparks et al.
1996).

This inner jet is measured to be one-sided on kiloparsec scales,
with the ratio of southern to northern jet fluxes greater than
about 3 for the soft X-ray band (Fig. 9) and greater than 4
from the radio data of Ledlow et al. (2002). This does
not challenge the apparent alignment of the large-scale radio
structure with the poles of the spiral disk, since the effects of
relativistic beaming are significant even close to the plane of
the sky. For a typical spectral index $\alpha=-0.7$ and using
the standard equation for jet-counterjet ratio $R$,
$$ R = \left( {{1 + \beta \cos \theta} \over 
{1 - \beta \cos \theta}}\right) ^{2 - \alpha}
\eqno{(1)} ,$$
$R>4$ and $\beta \approx 1$ allow any value of the angle between jets and
plane of the sky $\theta > 15^\circ$, comparable to the projected
angle between jets and poles of the galaxy disk. The observed jet asymmetry
is therefore consistent with the notion that the jets emerge within
$30^\circ$ of the polar direction, encountering minimal interstellar
gas along the way.

\section{Conclusions: Radio source production in a spiral host}

These results argue convincingly that 0313-192 is in fact a spiral galaxy
(albeit one which we see at a challenging orientation for morphological 
classification).
This prompts us to consider what factors might allow such a demonstrably rare
occurrence -- not only a kpc-scale radio jet, but an extensive and
powerful FR I double radio source -- beginning with the 
salient galaxy properties.

The host galaxy of the radio source 0313-192 is a large, bright spiral, roughly
Hubble type Sb. It has a luminous bulge, slightly more luminous than
that of M104. The extensive disk is twice
as bright as the bulge in the optical, and
we can trace luminous H II regions out to at least 17 kpc
from the nucleus. The dust lane is thick, richly structured
including features reaching several kpc from the midplane, and
is seen to nearly 20 kpc from the core. The dust lane shows a
warp of several degrees with respect to the stellar disk.

We see the nucleus through an amount of foreground material
typical of edge-on, gas-rich spiral disks, as seen both from
soft X-ray and 21-cm H I absorption. Likewise, the nucleus is
completely obscured in the optical, shortward of about 1.6 $\mu$m.
The emission lines seen in optical spectroscopy arise mostly
in a roughly linear structure seen emerging behind the dust lane
on both sides at an angle of about $20^\circ$; this structure
is seen nearly perpendicular to the radio jets.

If a tidal interaction has taken place, as might be
suggested by the disk warp and more
strongly inclined inner ionized-gas features, it must have
been at large mass ratio (such as acquisition of a gas-rich former
companion). The only potential companions close to 0313-192 are
dwarf systems. The apparently rich environment, in projection,
is part of Abell 428; this catalogued cluster shows such a large
velocity dispersion that it may be largely a large-scale filament
seen end-on rather than a traditional galaxy cluster. The redshift
data presented by Ledlow et al.\ (2001) indicate that this filament
is broken up into a number of small groups, which could well
have associated intragroup hot gas. The environment
is important, since production of large-scale radio sources is
fostered by interaction with an external medium. However, 
there is no specific evidence that the external gaseous environment
around 0313-192 is unusually dense for a spiral, since a large
fraction of spirals are found in groups.

We can picture several circumstances which may contribute to the
rarity of extensive double sources in spiral galaxies, and to the
ability of 0313-192 to produce such a source. The galaxy is
large and luminous, with a bulge luminosity suggesting a central
black-hole mass of order $8 \times 10^8$ $M_\sun$. This fits
with statistical evidence that radio-loud active nuclei have
an occurrence which is a strong function of the host
galaxy's stellar mass (Best et al. 2005). Furthermore,
the jet axis is projected roughly along the minor axis of the 
galaxy (at an angle of $22^\circ$). A polar orientation would
make jets more likely to escape the dense interstellar medium of
a spiral rather than having their energy dumped into the ISM
in a disordered way. Nearby Seyfert galaxies furnish several
examples of jets on small scales which lose their collimation rather
quickly through interaction with the interstellar medium. 
In Mkn 78, two small-scale jets are deflected or decollimated during
encounters which produce emission-line features (Whittle \& Wilson 2004).
The anomalous ``radio arms" of NGC 4258 are well described as
stalled, precessing jets, whose interactions with disk gas can be observed
in some detail (Cecil et al. 2000), providing a striking example of
how a normal galaxy's interstellar medium can dictate the fate of
jets as they propagate from the nucleus.
Since the black-hole masses
inferred for most Seyferts lie below the mass we suggest from the
bulge properties of 0313-192, its jets may well emerge with
a higher velocity than the ``weak, slow, and heavy" jets 
(Whittle et al. 2004) seen in many Seyferts.

A requirement for the formation of extensive radio lobes is the
escape of
jets to eventually encounter an intergalactic medium, which
would occur only if they avoided encountering high
column densities of cool interstellar gas, possible at high
angles to the plane in those spirals with bulges and nuclei that
are relatively free from such gas (a condition which probably excludes
the many Seyfert galaxies with surrounding star-forming regions,
for example).
 
Aside from 0313-192, the best-attested {\it disk} galaxy hosting
luminous radio jets and lobes is the S0 galaxy NGC 612 
(V{\'e}ron-Cetty \& V{\'e}ron 2001). Their analysis confirmed that
it is a bulge-dominated S0 system with stellar disk and dust annulus,
confirming the photographic results of Ekers et al.\ (1978). From
their photometric decomposition, the bulge-to-disk ratio in NGC 612 is
2.2 in the $i$ band, much smaller than the value of 0.4 we find from
modelling the images of 0313-192. In this comparison, the bulge in NGC 612 
remains over twice as luminous as we find for 0313-192.


Our original interest in studying 0313-192 was to understand the circumstances
which allow a spiral galaxy to produce a large double radio source with
attendant jets. This detailed analysis leads us to several clues. The
galaxy is large and luminous, with a particularly luminous bulge.
This in turn suggests that the central black hole is unusually
massive for a spiral, so that jets might emerge from its vicinity
faster than found in many Seyfert galaxies. The jets emerge roughly
along the poles of the disk; if this is needed for them to escape,
a further geometric factor limits the number of spirals which
could host such radio sources even during episodes of accretion
of the black hole. A final contributing factor in this case may
be a recent minor merger or weak tidal encounter, which would account for
the warped disk and perhaps for more strongly inclined gaseous structure
which we see within a few kpc of the nucleus.

\acknowledgments
This work was supported by NASA through STScI grant HST-GO-09376.01-A, and
through Chandra Award Number GO4-5118X issued by the Chandra
X-ray Observatory Center, which is operated by the Smithsonian Astrophysical
Observatory for and on behalf of the National Aeronautics Space Administration
under contract NAS8-03060. This research has made use of the USNOFS Image 
and Catalogue Archive
operated by the United States Naval Observatory, Flagstaff Station
(http://www.nofs.navy.mil/data/fchpix/). Fig.\ 1 attests to the Photoshop
wizardry of Zolt Levay. Chris Howk kindly provided a set of WIYN images
of edgewise spirals, of which NGC 4565 appears in Fig.\ 6. The NGC 891
image was taken, in collaboration with Gene Byrd, using the Nordic Optical 
Telescope, which is operated
on the island of La Palma jointly by Denmark, Finland, Iceland,
Norway, and Sweden, in the Spanish Observatorio del Roque de los
Muchachos of the Instituto de Astrofisica de Canarias. We thank
Gijs Verdoes Kleijn for interesting exchanges on how disk galaxies
might host strong radio jets, and Lisa Frattare for pointing us to
the NGC 5866 Hubble Heritage imagery.

\begin{figure}
\epsscale{0.3}
\caption
{The radio galaxy 0313-192 and its environment, in a color composite
produced from the ACS images in F555W and F814W. The red overlay
shows VLA 20-cm structure, encompassing a wide field of 
$82 \times 96$ arcseconds
to show most of the radio source. Multiple arrays have been
combined to retain higher resolution in the kpc-scale jet. This
image is available as STScI-PRC03-04. North is about 
$20^\circ$ counterclockwise from the top. 
The less inclined spiral has a
substantially different redshift and, while likely part of Abell 428,
does not form a bound interacting system with 0313-192.
}
\end{figure}

\begin{figure}
\epsscale{0.3}
\caption
{The radio galaxy 0313-192, in a color composite
produced from the ACS images in F555W and F814W. The radio galaxy
is shown here with an offset logarithmic intensity scale
to compress the image's dynamic range. North is $5^\circ$ degrees
clockwise from the top. This display, covering $20.0 \times 28.3$ arcseconds, 
emphasizes the complex absorption structure of the disk dust. In the
color scheme, blue corresponds to the F555W ($V$) image, red to F775W ($I$), and
green to their mean.
}
\end{figure}

\begin{figure}
\epsscale{0.3}
\caption
{H$\alpha$ and [O III] images with the narrowband ramp filters,
after continuum subtraction. Color gradients leave negative residuals
in parts of the dust lane. Each image has been rotated by $23.4^\circ$ to
align with the mean starlight plane, and spans $ 14.2 \times 22.0$ arcseconds.
North-northwest is at the top.
}
\end{figure}

\begin{figure}
\epsscale{0.3}
\caption
{$JHK$ images of 0313-192. from FLAMINGOS on {\it Gemini-S}.
Each panel spans $9.5 \times 32.3$ arcseconds and is shown with north at the top.
}
\end{figure}

\begin{figure}
\epsscale{0.3}
\caption
{Photometric profiles, in the F814W $I$ band, of the edge-on stellar
disk in 0313-192. The best-fit de Vaucouleurs bulge model was subtracted
before producing these profiles. The top panel shows profiles parallel to
the disk plane, averaged over regions from 0.8-1.6" north and south
of the mean midplane. The curve shows the projection of a truncated
exponential disk, as is seen in nearby edge-on spirals. The lower panel
illustrates the ``vertical" profile normal to the disk plane,
which is well modelled (away from the dust lane) by an exponential,
again as seen for local edge-on spirals.
}
\end{figure}

\begin{figure}
\epsscale{0.3}
\caption
{\footnotesize{Comparison of dust structures in 0313-192, from the ACS $V$ image
(center), with
nearby objects having well-defined Hubble types. Each is shown on an
intensity scale which is logarithmic after a small offset (1\%
of peak) to reduce
the appearance of sky noise. The filaments of dust perpendicular to
the plane are shared among all the spirals, and 0313-192 shows 
well-developed perpendicular structures of this kind. Radial truncation 
of the dust disk relative to the stellar disk is characteristic of
S0 systems, as exhibited only by NGC 5866 at the top. The NGC 4565 
image is from Howk \& Savage (1999); NGC 891 was observed with the
Nordic Optical Telescope, and the remaining comparison images are taken from
the HST archive.
}}
\end{figure}

\begin{figure}
\epsscale{0.3}
\includegraphics[scale=0.68,angle=-90]{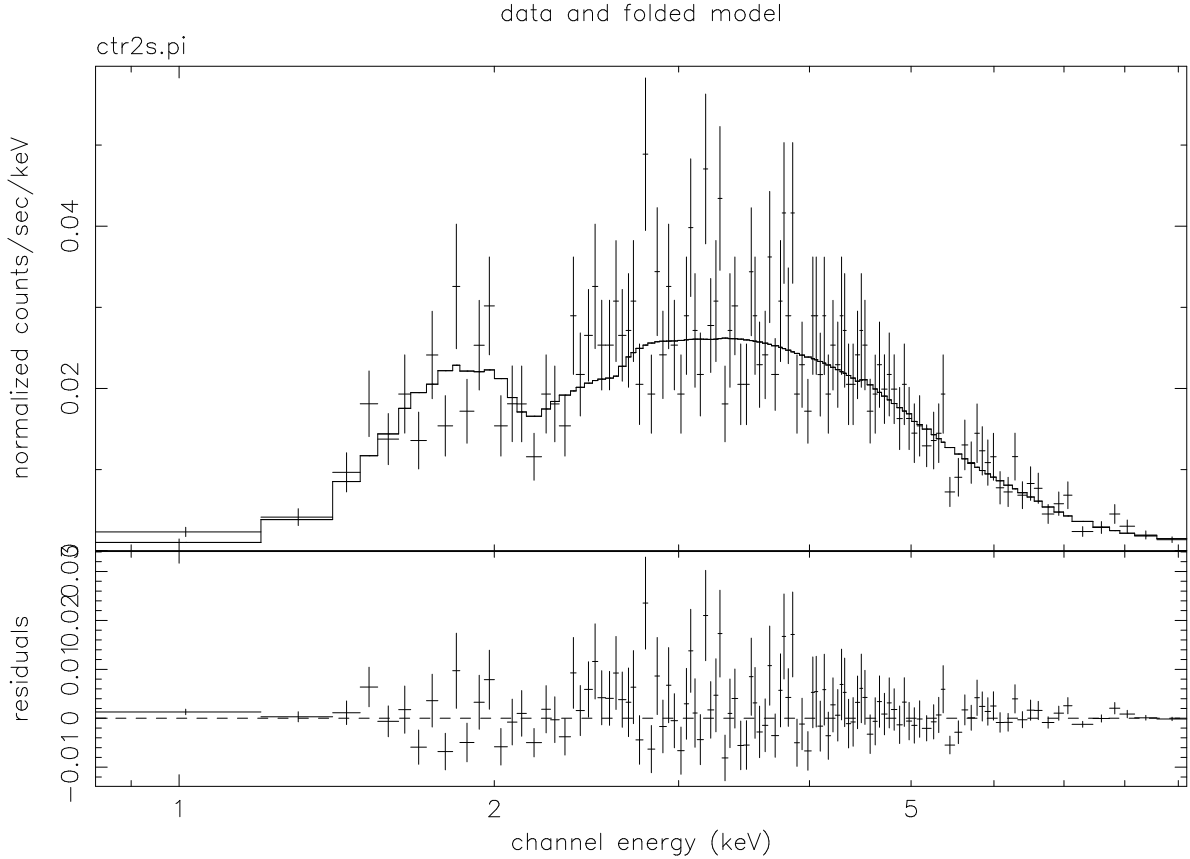}
\caption
{{\sl Chandra} ACIS spectrum of photons within 2\arcsec of the peak of 
0313-192, and the best-fit absorbed power-law fit. Properties of
this fit are: 
photon index $\Gamma=1.33^{+0.34}_{-0.27}$ and an intrinsic
absorbing column of $N_H=4.06^{+0.73}_{-0.58}\times10^{22}$ cm$^{-2}$
(90\% confidence intervals).
The lower
panel shows residuals from this fit.
}
\end{figure}

\begin{figure}
\epsscale{0.3}
\caption
{{\sl Chandra} ACIS images of the core region of 0313-192 in various
energy bands. At low energies, the strong absorption against the core
allows a small-scale X-ray jet to be seen (left). At intermediate
energies, we see a broader asymmetric halo of X-ray structure,
while the core source dominates at the highest energied measured (right).
Each image was produced by mapping the event list on a 0.125-\arcsec
grid, then convolving with a Gaussian of 0.75\arcsec FWHM.
}
\end{figure}

\begin{figure}
\epsscale{0.3}
\caption
{{\sl Chandra} ACIS image of the jet in 0313-192, using events from
0.35-1.2 keV and a smoothing of 1\arcsec FWHM, compared to the 
8.4 GHz structure as measured using the VLA (Ledlow et al. 2001). The
low-energy X-ray structure traces the radio jet.
}
\end{figure}

\end{document}